\documentclass[12pt,a4paper]{article}
\usepackage{jheppub}

\allowdisplaybreaks[4]
\usepackage{amsmath,bm}
\usepackage{subfigure}
\usepackage{pdfpages}
\usepackage{physics}
\usepackage{amsfonts}
\usepackage{mathtools}
\usepackage{color}

\newcommand{\kozuka}[1]{#1}
\newcommand{\murata}[1]{#1} \newcommand{\kinoshita}[1]{\if0{#1}\fi}

\title{
Quasinormal mode spectrum of the AdS black hole with the Robin boundary condition
}

\author{Shunichiro Kinoshita,}
\author{Tomohiro Kozuka,}
\author{Keiju Murata,}
\author{Keita Sugawara}
\affiliation{Department of Physics, College of Humanities and Sciences, Nihon University, Sakurajosui, Tokyo 156-8550, Japan}
\emailAdd{kinoshita.shunichiro@nihon-u.ac.jp}
\emailAdd{chto22004@g.nihon-u.ac.jp}
\emailAdd{murata.keiju@nihon-u.ac.jp}
\emailAdd{chkt23002@g.nihon-u.ac.jp}

\abstract{%
We study the quasinormal mode (QNM) spectrum of an asymptotically AdS black hole with the Robin boundary condition at infinity. We consider the Schwarzshild-AdS$_4$ with the flat event horizon as the background spacetime and study its scalar field perturbation. Denoting leading coefficients of slow- and fast-decay modes of the scalar field at infinity as $\phi_1$ and $\phi_2$, respectively, we assume a linear relation between them as $\phi_2 = \cot(\theta/2) \phi_1$, where $\theta$ is a constant called the Robin parameter and periodic under $\theta\sim\theta+2\pi$. In a certain range of the Robin parameter, there is an instability driven by the boundary condition. 
We also find the holonomy in the QNM spectrum under the parametric cycle of the boundary condition: $\theta=0\to2\pi$.
After the one-cycle,  $n$-th overtone of the QNM moves to $(n-1)$-th overtone. 
The fundamental tone of the QNM is swept out to the infinity in the complex plane.
}

\begin{document}
\maketitle

\section{Introduction}

Asymptotically AdS spacetimes have time-like boundary at infinity and are not globally hyperbolic.
To determine the dynamics of a field in the asymptotically AdS spacetime, 
it is necessary to specify a boundary condition at the boundary as well as the initial condition.
What boundary conditions are allowed at infinity? For concreteness, let us consider a massive free scalar field $\Phi$ satisfying 
$\Box \Phi=\mu^2\Phi$ in four dimensions.
When the metric asymptotes to that of the Poincare AdS$_4$ as $ds^2\simeq (-dt^2+dz^2+dx^2+dy^2)/z^2$ 
in the unit of AdS radius, the scalar field behaves as 
\begin{equation}
 \Phi(t,z,x,y)=z^{\Delta_-} (\phi_1(t,x,y) + \cdots) + z^{\Delta_+} (\phi_2(t,x,y) +  \cdots)\ \quad (z\sim 0),
 \label{1.1}
\end{equation}
where $\Delta_\pm = 3/2\pm \sqrt{9/4+\mu^2}$ and dots represent subleading terms determined by solving the equation of motion. We will call the first and second terms as slow- and fast-decay modes, respectively. 
One of the simplest choice of the boundary condition is $\phi_1=0$, i.e, there is no slow-decay mode.
However, this is not unique choice. In Refs.~\cite{Breitenlohner:1982jf,ishibashi:2004wx}, it has been shown that a finite energy can be defined even when $\phi_1\neq 0$ 
if the mass square is in 
\begin{equation}
 -\frac{9}{4}<\mu^2<-\frac{5}{4}\ .
\end{equation}
Then, we can consider the more general boundary condition as
\begin{equation}
 \phi_2 = \kappa \phi_1\ .
 \label{Robindef}
\end{equation}
For $\kappa=0$ and $\kappa=\pm\infty$, we simply have $\phi_2=0$ and $\phi_1=0$, respectively. We will call them as the Neumann and Dirichlet boundary conditions, respectively.
For $\kappa\neq 0, \pm\infty$, We will call the above condition as the Robin boundary condition. 

In the AdS/CFT correspondence~\cite{Maldacena:1997re,Gubser:1998bc,Witten:1998qj}, changing the boundary condition of the scalar field from Neumann to Robin corresponds to the double-trace deformation of the dual field theory~\cite{Witten:2001ua,Berkooz:2002ug,Mueck:2002gm,Minces:2002wp,Sever:2002fk,Gubser:2002vv,Aharony:2005sh,Elitzur:2005kz,Hartman:2006dy,Diaz:2007an,Papadimitriou:2007sj}. \kozuka{For the Neumann boundary condition, $\phi_1$ corresponds to the source $J$ coupled to the scalar operator $\mathcal{O}$ in the field theory. Then, the action of the field theory is deformed as $S\to S+\int J\mathcal{O}$. Under the Robin condition, the action is further deformed as $S\to S-\kappa\int \mathcal{O}^2/2+\int J\mathcal{O}$.} 
Thus, study of dynamics of the scalar field with the Robin boundary condition will be important for understanding the dynamics of the deformed theory.

Dynamics of the scalar field in asymptotically AdS spacetimes with the Robin boundary condition has been studied in several previous works.
In~Ref.~\cite{Araneda:2016ecy}, it was shown that there is an unstable mode for a sufficiently large negative value of $\kappa$ for general asymptotically AdS spacetimes.
In Refs.~\cite{Dappiaggi:2017pbe,Ferreira:2017tnc,Katagiri:2020mvm}, the effect of the Robin boundary condition on the superradiant instability of rotating and charged black holes in AdS was studied.
A nonlinear solitonic solution was also explicitly constructed under the Robin condition as a candidate of the final fate of the instability of the global AdS~\cite{Bizon:2020yqs}.  In Ref.~\cite{Harada:2023cfl}, hairy charged black holes and charged boson stars under the Robin conditions were constructed and their phase structure has been revealed.
Quasinormal modes of the Maxwell and Dirac perturbations in asymptotically AdS spacetimes have also been studied in Refs.~\cite{Wang:2021uix,Wang:2021upj,Wang:2019qja,Wang:2017fie,Wang:2015goa,Wang:2015fgp}.
\kozuka{For gravitational and Maxwell perturbations, as seen in previous works
\cite{Morgan:2009pn,Miranda:2005qx,Miranda:2008vb}, the notion of the Dirichlet, Neumann and Robin boundary conditions depends on 
which variable or which component one will choose as a perturbation field to impose the boundary condition.
For the scalar field perturbation, on the other hand, there is no ambiguity in the choice of the variable and also there is a clear meaning of the Robin parameter in the dual field theory.
}

In this paper, we will focus on the quasinormal mode spectrum of a scalar field in the Schwarzschild-AdS$_4$ spacetime with the Robin boundary condition\kozuka{, while in most previous studies, QNM spectra of scalar fields were studied with the Dirichlet boundary condition (for example, see \cite{PhysRevD.68.044024,PhysRevD.66.044009,PhysRevD.107.024023}). }
If a black hole is perturbed, the perturbation will be dissipated into the horizon. The relaxation process of the black hole perturbation is characterized by a complex-valued eigenfrequency called the quasinormal mode (QNM) frequency~\cite{Chandrasekhar:1975zza}.
(See~\cite{Kokkotas:1999bd,Berti:2009kk,Konoplya:2011qq} for reviews.) In the view of the AdS/CFT, 
certain asymptotically AdS black holes are considered as gravitational duals of thermal states in dual field theories and 
quasinormal mode spectra of AdS black holes have information about the equilibration of the thermal states~\cite{Horowitz:1999jd}. It has also been proposed that the QNM spectrum corresponds to poles of the retarded Green function of the thermal quantum field theory~\cite{Birmingham:2001pj,Son:2002sd,Kovtun:2005ev}. 
Especially, if there is an unstable mode in the gravitational theory, (i.e., if a QNM frequency is in the upper half of the complex plane), it indicates the instability of the thermal state and existence of a new stable phase in the dual field theory.

This paper is organized as follows. In section~\ref{toy}, we consider the ordinary wave equation in $(1+1)$ dimension with the Robin boundary condition as a toy example.
Studying the normal mode spectrum, we explicitly show that there is an unstable mode for a certain window of the Robin parameter $\kappa$.
In section~\ref{setup}, we study the QNM spectrum of the scalar field with the Robin boundary condition. We take the Schwarzschild-AdS$_4$ spacetime with flat horizon as the background.
Section~\ref{conc} is devoted to the conclusion. In appendix, the detail of the numerical computation of the QNM is summarized.

\section{Toy example: ($1+1$)-d wave equation with Robin boundary condition}
\label{toy}

\subsection{Spectrum}

To understand the effect of the Robin boundary condition, we consider the wave equation in ($1+1$)-dimensional flat spacetime: 
\begin{equation}
 \ddot{\phi}-\phi''=0\ ,
\label{waveeq}
\end{equation}
where ${}^\cdot\equiv \partial_t$ and ${}'\equiv \partial_x$.
We take the domain of the wave equation as $0\leq x \leq 1$.
We impose the Robin boundary condition at $x=0$ and the Dirichlet one at $x=1$ as 
\begin{align}
&\phi'|_{x=0} = \kappa \phi|_{x=0}\ ,\label{x0bc}\\
&\phi|_{x=1}=0\ ,\label{xLbc}
\end{align}
where $\kappa \in \mathbb{R}$ represents the Robin parameter: $\kappa=0$ and $\kappa=\pm\infty$ correspond to the Neumann and  Dirichlet boundary conditions, respectively.
\kozuka{ Since $\kappa=+\infty$ and $\kappa=-\infty$ provide the same Dirichlet condition, the parameter space of $\kappa$ can be identified with $\mathbb{S}^1$.} %
\footnote{
\kozuka{Precisely speaking, the boundary condition (\ref{x0bc}) can be rewritten as $(a\phi + b\phi')|_{x=0} = 0$ such that $\{(a:b)|(a,b)\neq(0,0)\}$, i.e., the one-dimensional real projective space $\mathbb{RP}^1$. This parameter space is topologically equivalent to $\mathbb{S}^1$. }
}

\begin{figure}
\centering
\includegraphics[scale=0.5]{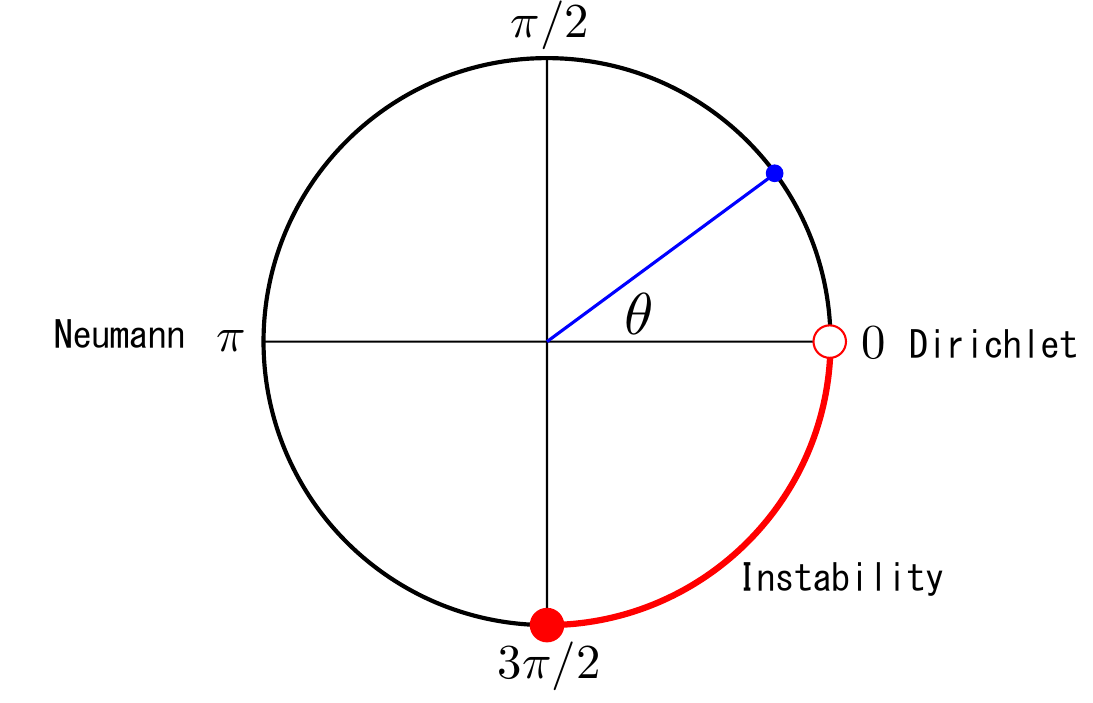}
\caption{Visualization of parameter space for the Robin boundary conditions ($\kappa = \cot(\theta/2)$). The Dirichlet and Neumann boundary conditions are given by $\theta =0$ and $\theta=\pi$, respectively.
On the portion of the circle shown by red, the trivial solution $\phi=0$ is unstable.
}
\label{bccircle}
\end{figure}

For the following analysis, it is convenient to introduce a parameter $\theta$ as
\begin{equation}
 \kappa=\cot \frac{\theta}{2}\ .
 \label{rp}
\end{equation}
In terms of $\theta$, the Dirichlet and Neumann boundary conditions correspond to $\theta=0$ and $\theta=\pi$.
The Robin boundary condition are specified by a point on the circle as in Fig.~\ref{bccircle}.
For any value of $\theta$, we have a trivial solution $\phi=0$. We will study the spectrum of the scalar field $\phi$ with conditions~(\ref{x0bc}) and (\ref{xLbc}).
In \cite{Bizon:2020yqs}, it has been shown that the Klein-Gordon equation for spherically symmetric scalar field in the global AdS$_4$ reduces to Eq.~(\ref{waveeq}) and instability has been found for certain Robin boundary conditions.\footnote{
\kozuka{In \cite{Bizon:2020yqs}, $\Phi(x)=\cot(x)\Psi(x)$ leads to Eq.~(\ref{waveeq}).}}
(See also ~\cite{Araneda:2016ecy} for the case of semi-infinite domain, $0\leq x < \infty$.)
\kozuka{ The novelty of our study in this section is to reveal how the spectrum of the normal modes moves on the complex plane by varying the \murata{Robin parameter} $\theta$, as well as to explain the instability. We will explicitly confirm the existence of holonomy of the normal mode spectrum by the parametric cycle $\theta\to \theta + 2\pi$.
This study will help us to understand results of quasinormal modes with the Robin boundary condition in subsequent sections.}

\begin{figure}
  \centering
  \subfigure[$\omega \in \mathbb{R}$]
 {\includegraphics[scale=0.7]{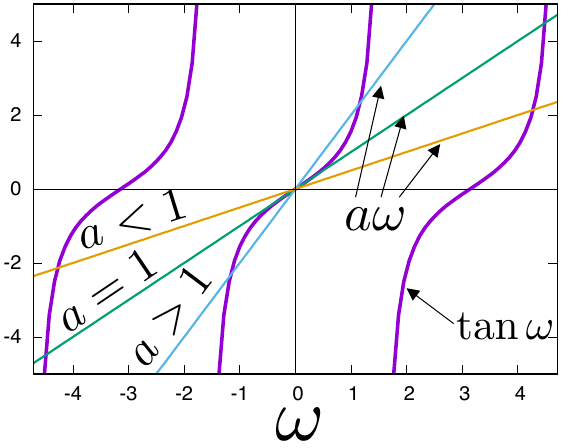}\label{tan}
  }
  \subfigure[$\omega \in i\mathbb{R}$ \quad ($\omega\equiv i\Omega$)]
 {\includegraphics[scale=0.69]{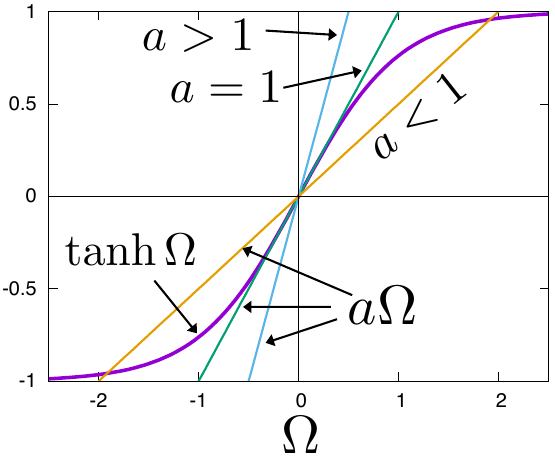}\label{tanh}
  }
  \caption{Illustration for determining spectrum. 
}
  \label{tantanh}
\end{figure}

\begin{figure}
\centering
\includegraphics[scale=0.5]{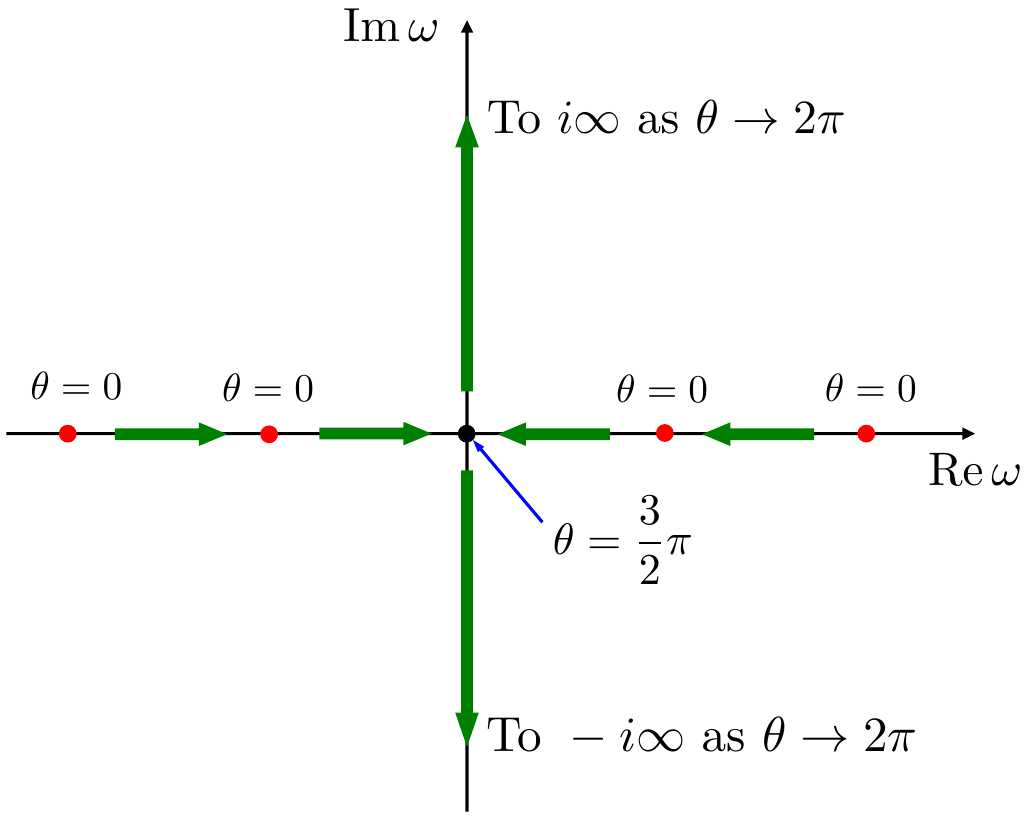}
\caption{Motion of eigenfrequencies in the complex $\omega$-plane.
}
\label{complexomega}
\end{figure}

Let us consider a single mode whose frequency is $\omega$. By only imposing the boundary condition~(\ref{xLbc}) at $x=1$, we have a solution of the wave equation as
\begin{equation}
 \phi (t,x)= e^{-i\omega t}\sin \omega (1-x)\ .
\end{equation}
Substituting the above expression into the Robin boundary condition~(\ref{x0bc}) at $x=0$, we obtain an equation which determines the spectrum: 
\begin{equation}
 a \omega   = \tan \omega\  ,\qquad (a\equiv -\tan\frac{\theta}{2})\ .
\label{taneq}
\end{equation}
For the Dirichlet ($\theta=0$) and Neumann ($\theta=\pi$) conditions, 
the spectra are given by $\omega_n = n \pi$ $(n\in \mathbb{Z}\backslash\{0\})$ and $\omega_n = (n+1/2)\pi$ ($n\in \mathbb{Z}$), respectively. 
Although, for a general value of $\theta$, we cannot obtain analytical solutions, we can see the $\theta$-dependence of the spectrum graphically~\cite{Bizon:2020yqs}. 
Figure~\ref{tan} shows each side of Eq.~(\ref{taneq}) as functions of $\omega \in \mathbb{R}$. The intersections correspond to the spectrum given by Eq.~(\ref{taneq}).
In the region of $|\omega| < \pi/2$, the number of intersections depends on the slope of the linear function, $a= -\tan(\theta/2)$. 
Imagine that we start from $a > 1$ and decrease the value of $a$. Initially, there are three intersections in $|\omega | < \pi/2$. 
For $a =1$, however, all the three intersections merge at $\omega =0$. For $a < 1$, we only have single intersection at the origin.
The critical case $a=1$ corresponds to $\theta=3\pi/2$.
Where have lost two solutions gone to? They have gone to the complex plane. 
To see this, we assume that the frequency $\omega$ is pure imaginary and denote $\omega=i\Omega$. Then, Eq.~(\ref{taneq}) is rewritten as
\begin{equation}
 a \Omega = \tanh \Omega \ .
\label{tanheq}
\end{equation}
Figure~\ref{tanh} shows each side of Eq.~(\ref{tanheq}) as functions of $\Omega$. 
For $a < 1$, we can find two pure imaginary solutions. The mode with $\textrm{Im}\,\omega>0$ grows exponentially in $t$ and implies instability of the trivial solution $\phi=0$.
Figure~\ref{complexomega} shows the motion of eigenfrequencies on the complex $\omega$-plane when we change the value of $\theta$ from $0$ to $2\pi$.
For the Dirichlet condition $\theta=0$, all eigenfrequencies are located at the red points on the real axis. For the critical case $\theta=3\pi/2$, two of the eigenfrequencies merge at the origin. 
For $\theta>3\pi/2$, they split into the the imaginary axis and the unstable mode appears. For the Dirichlet condition after the one-cycle,  $\theta=2\pi$, the two eigenfrequencies on the imaginary axis 
swept out to the infinity  and this is consistent with the stability of the trivial solution for the Dirichlet condition.
By the one-cycle $\theta=0 \to 2\pi$, normal mode frequencies move as $\omega_{n+1}\to \omega_n$ $(n\geq 1)$ and $\omega_{n-1}\to \omega_n$ $(n\leq -1)$. The fundamental mode frequencies go to infinity 
$\omega_{\pm 1}\to \pm \infty i$ for $\theta\to 2\pi$. 
It follows that we have the same distributions of normal mode frequencies for $\theta=2\pi m$ ($m\in \mathbb{Z}$).
When the eigenfrequency is pure imaginary $\omega=i\Omega$, the mode function becomes
\begin{equation}
    \phi = i e^{\Omega t}\sinh \Omega (1-x)\ .
\end{equation}
This function is localized at $x\sim 0$ for a sufficiently large $\Omega$. 
In the limit $\Omega\to \pm \infty$, this becomes singular eventually.

For the critical case $\theta=3\pi/2$, we can explicitly find the solution of the wave equation:
\begin{equation}
 \phi(t,x)=Ct(1-x)\ ,
\end{equation}
where $C$ is an arbitrary constant. This grows linearly in $t$. Thus, $\theta=3\pi/2$ is marginally unstable. 
In summary, the condition for instability of the trivial solution $\phi=0$ with the Robin boundary is given by
\begin{equation}
 \frac{3}{2}{\pi} \leq \theta < 2\pi\ 
 \quad
 \left(\kappa \leq -1\right) .
\label{instcond}
\end{equation}
The unstable boundary conditions are illustrated in Fig.~\ref{bccircle}.

\subsection{Origin of instability}
\label{subsec:origin_flat}

We have seen that the ordinary wave equation is unstable when the inequality~(\ref{instcond}) is satisfied.
The unstable mode grows exponentially and should have a larger energy at the late time comparing to that of the initial state.
Where did the energy come from? We give an energetics of the instability in this subsection.

We can naively define the Hamiltonian of the current system as
\begin{equation}
 H = \frac{1}{2}\int_0^1 \!\! dx \, [\dot{\phi}^2(t,x) + \phi'^2(t,x)] .
\end{equation}
Its time derivative is  
\begin{equation}
 \begin{aligned}
  \frac{dH}{dt} &= 
  \int_0^1 \!\! dx \, [\dot{\phi}(t,x)\ddot{\phi}(t,x)
  + \phi'(t,x)\dot\phi'(t,x)] \\
  &= \int_0^1 \!\! dx \, \frac{\partial}{\partial x}
  [\dot{\phi}(t,x)\phi'(t,x)] \\
  &= - \dot{\phi}(t,0)\phi'(t,0) ,
 \end{aligned}
\end{equation}
where we have imposed the Dirichlet condition $\phi=0$ at the other side $x=1$.
This means that the energy cannot be always conserved due to a flux injected from the boundary 
depending on the boundary condition at $x=0$.
Now, at $x=0$, we consider the Robin boundary condition:  
$\phi(t,0) \cos(\theta/2)  - \phi'(t,0) \sin(\theta/2) = 0$.
Then, we have 
\begin{equation}
  \frac{dH}{dt} = - \cot\frac{\theta}{2} \, \phi(t,0)\dot\phi(t,0) 
  = - \frac{d}{dt}\left(\frac{\cot(\theta/2)}{2}\phi^2(t,0)\right)\ .
\end{equation}
Thus, it turns out that the ``modified'' Hamiltonian 
\begin{equation}
\tilde{H}=\frac{1}{2}\int_0^1 \!\! dx \, [\dot{\phi}^2(t,x) + \phi'^2(t,x)] + \frac{1}{2}\phi^2(t,0) \cot(\theta/2)
\label{modH}
\end{equation}
is conserved under the Robin condition. The integral of the above expression is clearly non-negative.
On the other hand, the last term can be negative depending on the sign of $\kappa=\cot(\theta/2)$.
\kozuka{For the Robin boundary condition with $\kappa<0$, the negative energy can be stored at the \kinoshita{AdS} boundary and the energy of the bulk is not bounded. 
Thus, $\phi(t,x)$ would be unbounded and this system can be unstable. The negative energy at the \kinoshita{AdS} boundary can be regarded as an origin of the instability.}

Let us now carefully consider the boundary term of the action. 
An action that provides the wave equation (\ref{waveeq}) is simply given by 
\begin{equation}
 S[\phi] = \int^{t_2}_{t_1} \!\! dt \int_0^1 \!\! dx \, \frac{1}{2}
  [\dot{\phi}^2(t,x) - \phi'^2(t,x)] .
\end{equation}
Its first variation for $\phi + \delta\phi$ becomes 
\begin{equation}
 \begin{aligned}
  \delta S &= \int^{t_2}_{t_1} \!\! dt \int_0^1 \!\! dx \, 
  [\dot{\phi} \delta\dot\phi - \phi'\delta\phi'] \\
  &= \int^{t_2}_{t_1} \!\! dt \, \phi'(t,0) \delta\phi(t,0) ,
 \end{aligned} 
 \label{eq:deltaS}
\end{equation}
where we have used the equation of motion. 
The first variation of the original action $S$ 
does not always vanish 
with respect to an arbitrary variation $\delta\phi$ at $x=0$.
We define an action with a total derivative term as 
\begin{equation}
 \begin{aligned}
  \widetilde{S} &\equiv S - \frac{\cot(\theta/2)}{2} \int^{t_2}_{t_1} \!\! dt \, \phi^2(t,0)
  \\
  &= S + \cot\frac{\theta}{2} 
  \int^{t_2}_{t_1} \!\! dt \int_0^1 \!\! dx \, \phi(t,x)\phi'(t,x) .
 \end{aligned}
 \label{eq:mod_action}
\end{equation}
We note that this modified action $\widetilde{S}$ can satisfy $\delta \widetilde{S} = 0$ 
with respect to an arbitrary variation $\delta\phi$ at $x=0$ 
imposing the Robin boundary condition.
We consider the invariance of the action in terms of time translation $t \to t + \epsilon$.
The Noether theorem states that a conserved charge exists on shell.
What we emphasize is that it relies on use of the boundary conditions as well as the equations of motion. 
Under the Robin boundary condition, the variation of the modified action \kozuka{yields no} boundary contribution, so that the Noether charge derived from $\widetilde{S}$ can be conserved without any flux on the boundary.
 
From the modified Lagrangian density: 
\begin{equation}
 \widetilde{\mathcal{L}} = \frac{1}{2}[\dot{\phi}^2 - \phi'^2]
  + \cot\frac{\theta}{2} \, \phi\phi' ,
\end{equation}
we have  
the Hamiltonian density:  
\begin{equation}
 \widetilde{\mathcal{H}} = \frac{1}{2}[\pi^2 + \phi'^2]
  - \cot\frac{\theta}{2} \, \phi\phi' 
\quad \left(\pi \equiv \frac{\partial\widetilde{\mathcal{L}}}{\partial\dot{\phi}}
      = \dot\phi\right) .
\end{equation}
This provides a conserved energy in this system as 
\begin{equation}
       \tilde{H} = \int^1_0 dx \widetilde{\mathcal{H}} 
       = \frac{1}{2}\int^1_0 dx [\dot\phi^2 + \phi'^2]
       + \frac{1}{2}\cot\frac{\theta}{2} \phi(t,0)^2 .
\end{equation}
Thus, we can reproduce Eq.~(\ref{modH}) as the conserved quantity again.

In fact, the boundary condition at $x=0$ locally causes the two pure imaginary modes that we have seen previously.
To understand it easily, let us consider the following one-dimensional Sch\"odinger problem: 
\begin{equation}
 \left[- \frac{d^2}{dx^2} + 2\kappa \delta(x)\right] \phi_\omega(x) 
  = \omega^2 \phi_\omega(x) ,
\end{equation}
\kozuka{where $\delta(x)$ is the Dirac delta function \murata{and $\phi_\omega(x)$ denotes a Fourier component given by $\phi(t,x)=\int^\infty_{-\infty} d\omega \phi_\omega(x) e^{-i\omega t}$}.}
Integrating it within $-\epsilon \le x \le \epsilon$, we have  
\begin{equation}
 - \left.\frac{d}{dx}\phi_\omega(x)\right|^{\epsilon}_{-\epsilon}
  + 2\kappa \phi_\omega(0) 
= 0 ,
\end{equation}
which provides the Robin boundary condition, 
$\phi'_\omega(0)=\kappa \phi_\omega(0)$, at $x=0$ \kozuka{ in the limit $\epsilon \to 0$}.
This means that imposing the Robin boundary condition at $x=0$ is equivalent \kozuka{to adding the} delta-function potential at $x=0$, whose coefficient is determined by $\kappa$.

Now, we will focus on local behavior near $x=0$ irrelevant to boundary condition at the other side.
Imposing outgoing boundary conditions $\phi_\omega(x) \sim \exp (+ i\omega x)$ for $x \to \infty$
and $\phi_\omega(x) \sim \exp (- i\omega x)$ for $x \to - \infty$,
a solution is 
\begin{equation}
 \phi_\omega(x) = c \exp (\kappa |x|) ,\quad \omega = \pm i\kappa .
\end{equation}
If $\kappa < 0$, this solution describes a bound state with 
$\omega^2 <0$ by the delta potential well.
\kozuka{ This implies $\phi(t,x) \sim \exp(-i\omega t)\phi_\omega(x)$ will exponentially grow \murata{along the time-like Killing vector $\partial_t$}.}
It turns out that the existence of the ``bound state with delta potential'' is origin of the instability 
under the Robin boundary condition with negative Robin parameters.

\section{QNM spectrum of the scalar field with the Robin boundary condition}\label{setup}
\subsection{Setup}

We consider the four-dimensional Schwarzschild-AdS spacetime (Sch-AdS$_4$) with the flat horizon as the background spacetime,
\begin{align}
    ds^2
    =\frac{l^2}{z^2}\qty(-F(z)dt^2+\frac{1}{F(z)}dz^2+dx^2+dy^2)\ ,\quad F(z)=1-\frac{z^3}{z_h^3} ,
    \label{SchAdS4wfh}
\end{align}
where $l$ is the AdS radius and the event horizon is located at $z=z_h$.
This metric satisfies the Einstein equation in vacuum: $G_{\mu\nu}-3l^{-2}{g_{\mu\nu}}=0$.
\kozuka{ In the case of the plane symmetric horizon, we can set $z_h=1$ by the scaling transformation $x^\mu\rightarrow z_h x^\mu \ (x^\mu=t,x,y,z)$. We can also choose units in which $l=1$. Thus, from now on, we can consider the metric \eqref{SchAdS4wfh} only for $l=1$ and $z_h=1$, without loss of generality. }
In the ingoing Eddington-Finkelstein (EF) coordinates $(v,z)$, the metric 
(\ref{SchAdS4wfh}) can be rewritten as
\begin{equation}
    ds^2=\frac{1}{z^2}(-F(z)dv^2-2dvdz+dx^2+dy^2) ,
    \label{iEFcSchAdS4wfh}
\end{equation}
where
\begin{equation}
    v=t+r_*,\quad r_*=-\int^{z}_{0}\frac{d\Tilde{z}}{F(\Tilde{z})} .
    \label{iEFc}
\end{equation}
Note that the tortoise coordinate $r_*$ behaves as $r_* \simeq -z$ and $r_* \to -\infty$ near the AdS boundary ($z \sim 0$) and the horizon ($z=1$), respectively. 

We consider the Klein-Gordon equation 
\begin{equation}
    \Box\Phi-\mu^2\Phi=0 ,
    \label{K-G-eq}
\end{equation}
in the Sch-AdS$_4$ spacetime.
The scalar field $\Phi$ can be expanded by Fourier modes as
\begin{equation}
    \Phi(v,z,x,y)=\varphi(z)e^{-i\omega v+i\kozuka{\bm{k}\cdot\bm{x}}}=\phi(z)e^{-i\omega t+i\kozuka{\bm{k}\cdot\bm{x}}} ,
    \label{wavefunction}
\end{equation}
where $\varphi(z)$ and $\phi(z)$ are Fourier coefficients in $(v,z)$- and $(t,z)$-coordinates, respectively. \kozuka{We also define $\bm{k}=(k_x,k_y)$ as $2$-dimensional wave number and $\bm{x}=(x,y)$ as $2$-dimensional coordinate vector.}
\murata{Since we do not assume that $(x,y)$-space is compactified, the wave number $\bm{k}$ is continuous and defined in $\mathbb{R}^2$. In cases of compactified $(x,y)$-space, it should be simply discretized.}
Note that, since $\phi(z) = \varphi(z)e^{-i\omega r_\ast}$, we have 
\begin{equation}
    \phi(z) \simeq \varphi(z)(1+i\omega z) \quad (z\simeq 0) .
    \label{varphiandohi}
\end{equation}
Substituting Eq.~(\ref{wavefunction}) into the Klein-Gordon equation {\eqref{K-G-eq}}, we obtain  the equation for $\varphi(z)$ as
\begin{equation}
    -\mu^2\varphi+z\left[-(k^2z+2i\omega)\varphi+(2iz\omega-2F+zF')\varphi'+zF\varphi''\right]=0 ,
    \label{EFKG}
\end{equation}
where $'\equiv d/dz$ \kozuka{and $k=|\bm{k}|$}. 
\kozuka{
The above equation can also be written as 
\begin{equation}
    \left[-\frac{d^2}{dr_\ast^2}+V(z)\right]\chi=\omega^2\chi\ ,\quad V(z)=\frac{F(z)}{z^2}(k^2 z^2 - zF'(z) + 2F(z) + \mu^2)\ ,
    \label{Schro}
\end{equation}
where $\chi(z)\equiv \phi(z)/z=\varphi(z)e^{-i\omega r_\ast}/z$. This is a Schr\"{o}dinger-like equation. 
}

Now, for simplicity, we consider that the mass square is
\begin{equation}
 \mu^2=-2\ .
 \label{mssq2}
\end{equation}
\kozuka{This describes the conformally coupled scalar field.}
In this case, the asymptotic form of the solution of the equation (\ref{EFKG}) at $z\sim0$ is given by 
\begin{equation}
    \varphi(z)\sim \varphi_1 z+\varphi_2z^2\ ,\qquad \phi(z)\sim \phi_1 z+\phi_2 z^2 .
    \label{asymphi} 
\end{equation}
\kozuka{Powers of asymptotic expansion of the scalar field are now integers and we can easily impose the Robin boundary condition at $z=0$ in our numerical calculations.
For a general mass square, we need to treat irrational power-law decay near the infinity for the scalar field.}

Then, from Eq.~(\ref{varphiandohi}), 
the above coefficients are related to each other as follows: 
\begin{equation}
    \phi_1=\varphi_1\ ,\quad \phi_2=\varphi_2+i\omega \varphi_1 .
\end{equation}
Imposing the Robin boundary condition~(\ref{Robindef}) in the $(t,z)$-coordinates, we obtain
\begin{equation}
    \varphi_2+i\omega \varphi_1 = \kappa \varphi_1 ,
    \label{varphiRobin}
\end{equation}
in the ingoing EF coordinates.
For later convenience, we introduce the rescaled  
variable as
\begin{equation}
    f(z)\equiv\frac{\varphi(z)}{z}\sim \varphi_1+\varphi_2z .
    \label{fphi}
\end{equation}
From Eqs.~(\ref{EFKG}), (\ref{mssq2}), and (\ref{fphi}), 
in terms of $f(z)$ the field equation can be rewritten as 
\begin{equation}
    -(k^2+z)f(z)+(2\lambda-3z^2)f'(z)+(1-z^3)f''(z)=0 ,
    \label{eqn0}
\end{equation}
where $\omega=-i\lambda$.
The Robin boundary condition~(\ref{varphiRobin}) is also written as
\begin{equation}
    f'(0)+(\lambda-\kappa)f(0)=0 .
    \label{boundary condition}
\end{equation}

\kozuka{Near the horizon, general solutions of the scalar field behave as $\Phi\sim e^{-i\omega t+i\bm{k}\cdot\bm{x}} (c_1 e^{-i\omega r_\ast}+c_2 e^{i\omega r_\ast})= e^{-i\omega v+i\bm{k}\cdot\bm{x}} (c_1 +c_2 e^{2i\omega r_\ast})$, where $c_1$ and $c_2$ are constants of integration.
In order to obtain QNMs, we should impose the the ingoing boundary condition corresponding to $c_2=0$. Therefore, in the ingoing EF coordinates we just have to require $f(z)$ to be regular at the future horizon $z=1$. 
}
If we obtain $\lambda$ and $f(z)$ that satisfy the above equations, the QNM spectrum and mode functions can be determined.

In later calculations, we mainly use the periodic parameter $\theta$ defined by equation (\ref{rp}) instead of $\kappa$ as the Robin parameter.
We can take its domain as $0 \leq \theta < 2\pi$ without loss of generality. 
Note that $\theta=0$ ($\kappa=\pm\infty$) corresponds to the Dirichlet boundary condition, while $\theta=\pi$ ($\kappa=0$) corresponds to the Neumann boundary condition. 

\subsection{The results of QNM spectrum and mode functions}

\begin{figure}
  \centering
  \subfigure[QNM spectrum for $\theta=0$]
 {\includegraphics[scale=0.21]{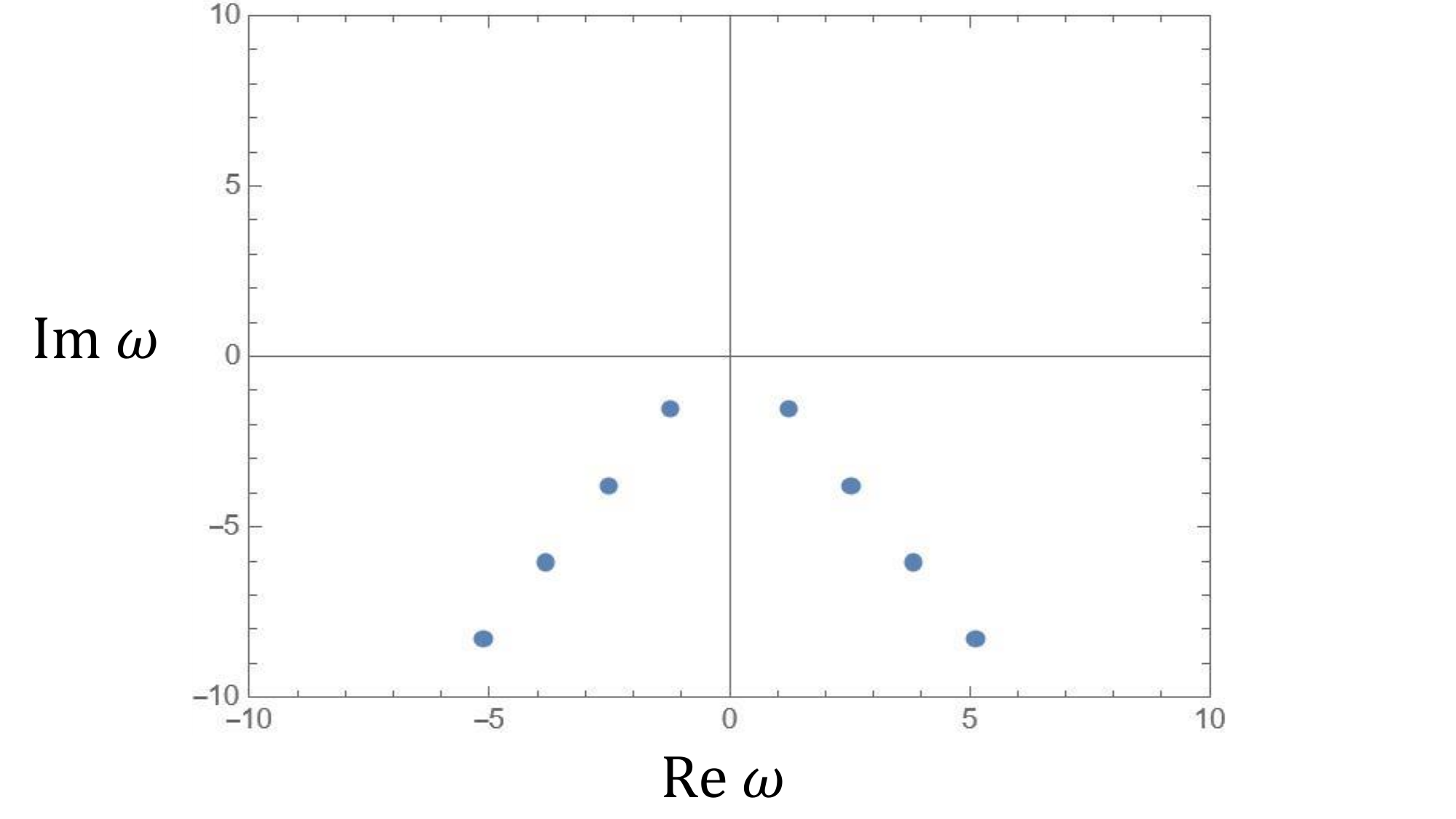}\label{QMMD}
  }
  \subfigure[QNM spectrum for $\theta=\pi$]
 {\includegraphics[scale=0.21]{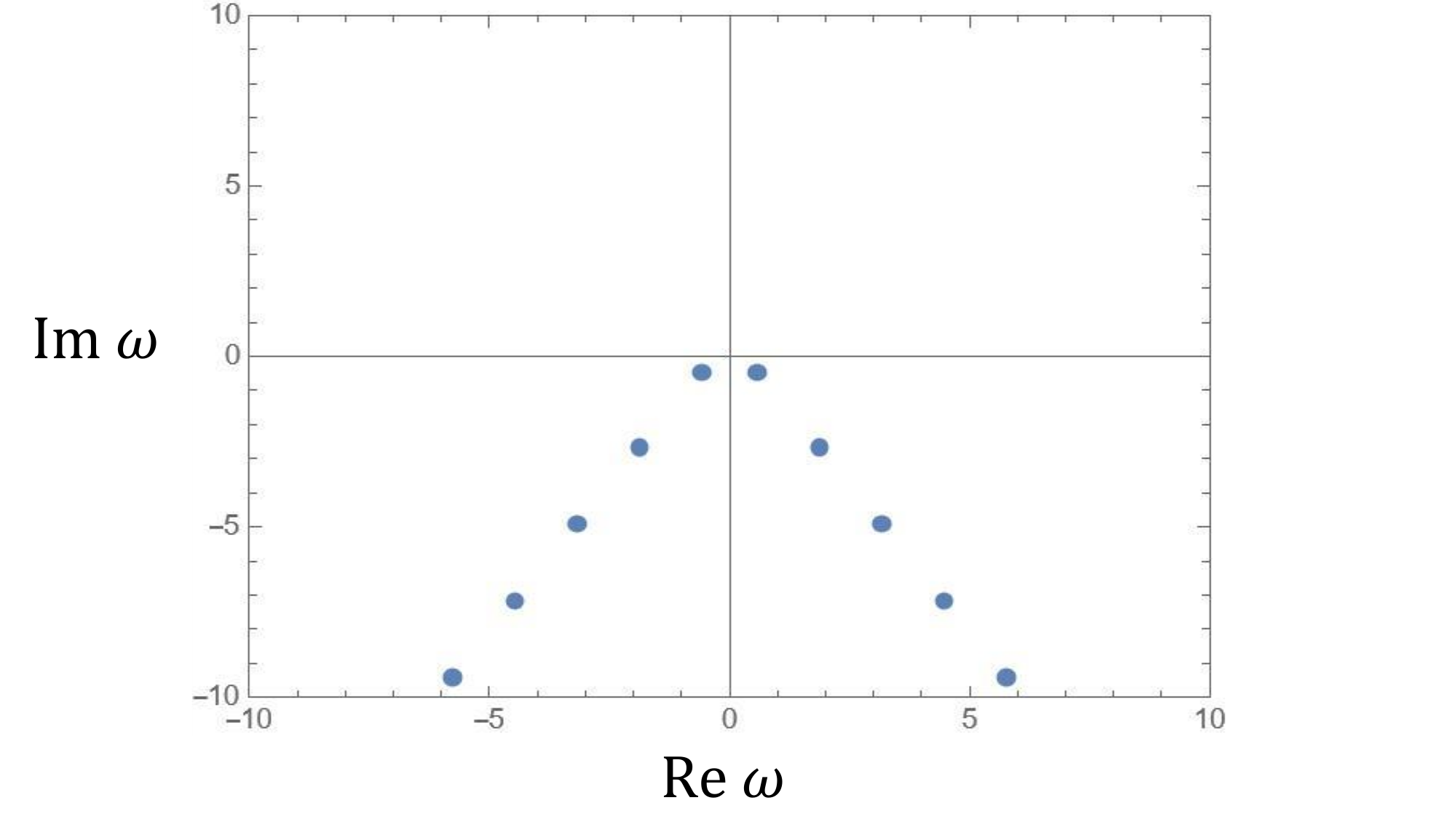}\label{QMMN}
  }
  \subfigure[QNM spectrum for $\theta=7\pi/4$]
 {\includegraphics[scale=0.21]{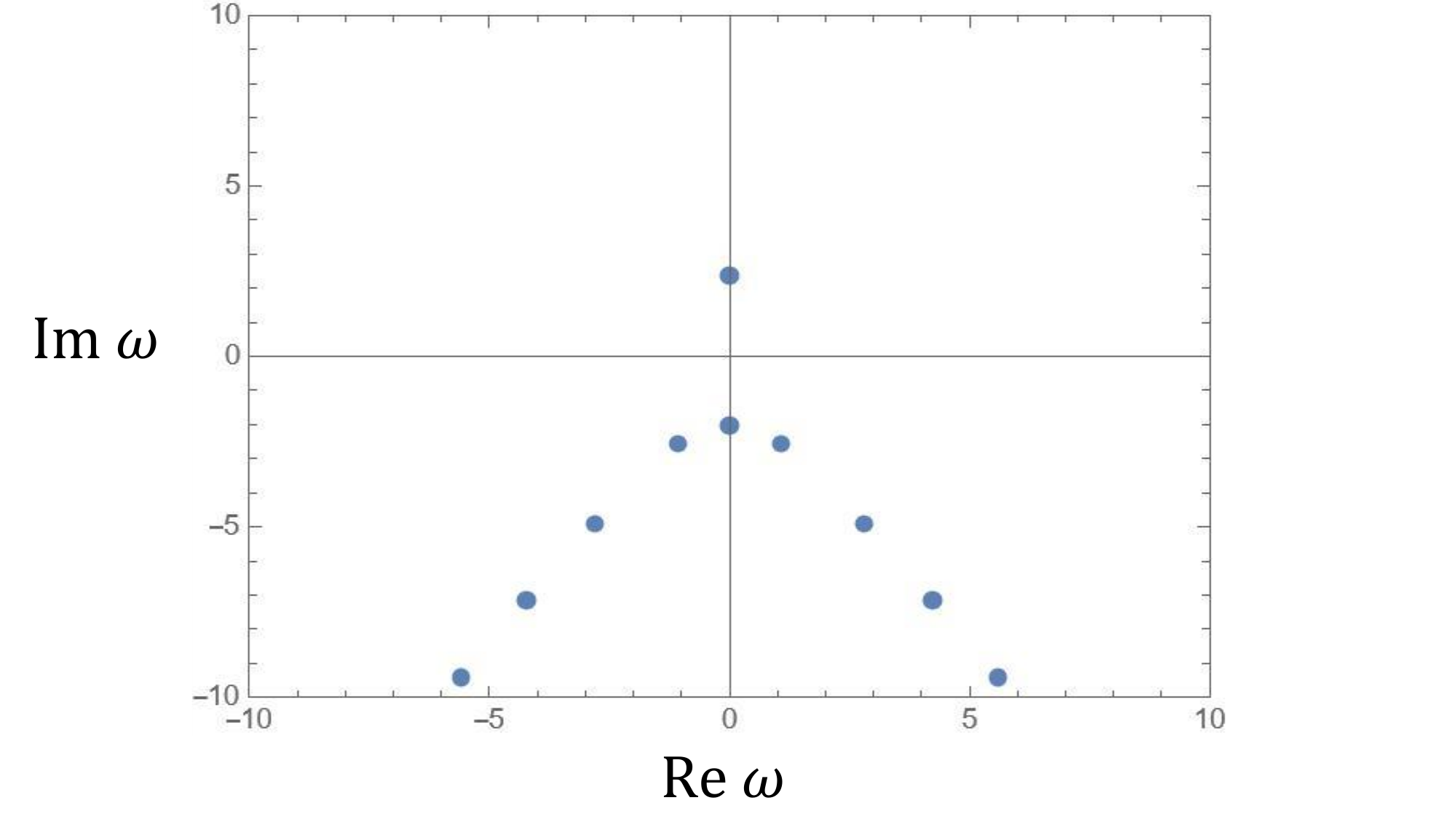}\label{QMMR}
  }
 \caption{
 (a)-(c) QNM spectra for $0$ (Dirichlet), $\theta=\pi$ (Neumann), $7\pi/4$ (Robin).
 }
\end{figure}

\begin{figure}[t]
    \centering
    \subfigure[$k=1$]{\includegraphics[scale=0.5]{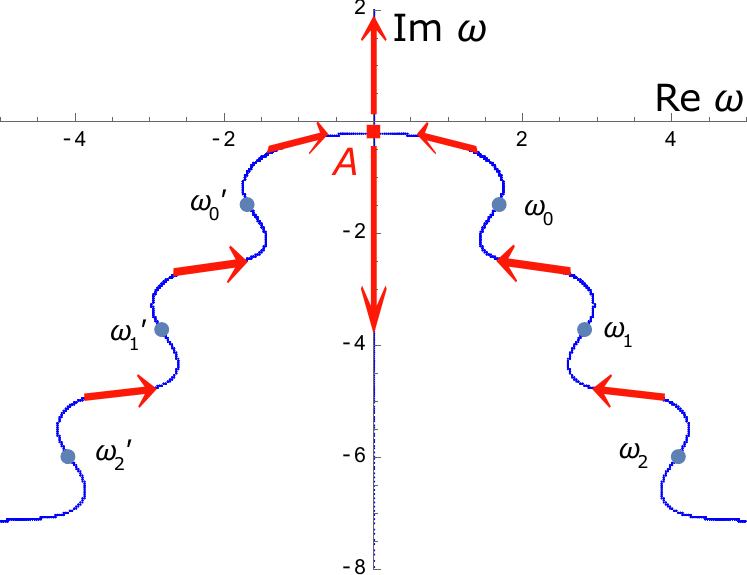}\label{QNMOrbits1}}
    \subfigure[$k=2$]{\includegraphics[scale=0.5]{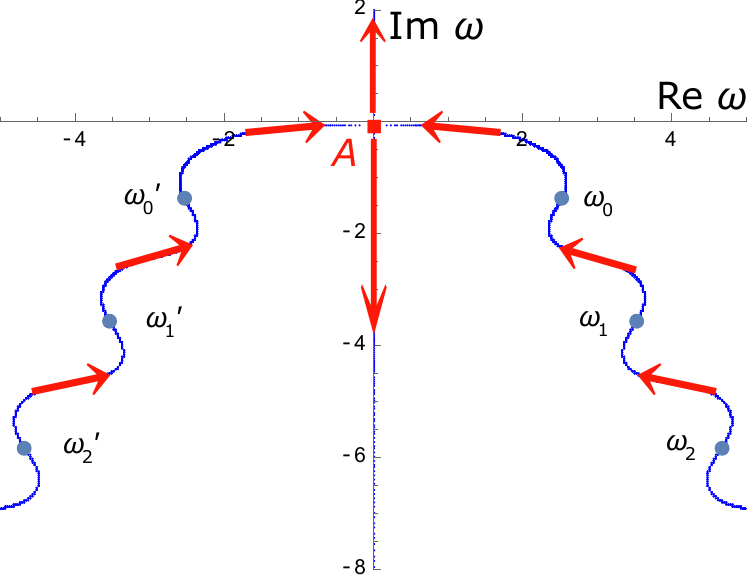}\label{QNMOrbits2}}
    \subfigure[$k=10$]{\includegraphics[scale=0.5]{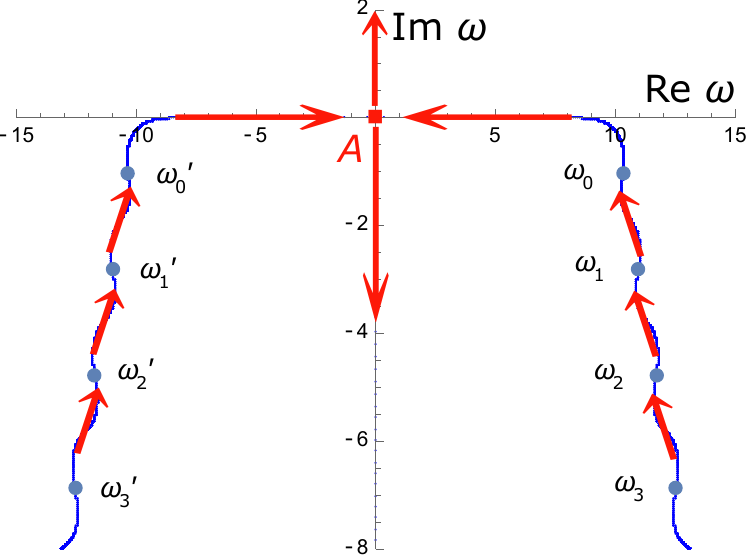}\label{QNMOrbits10}}
    \caption{The trajectory of quasinormal frequencies of wavenumber $k=1,2,10$ with respect to the Robin parameter $\theta$.}
    \label{QNMOrbits1210}
\end{figure}

In this subsection, we will show our numerical results on QNM spectra with the Robin boundary condition. We have used the spectral method for the numerical calculations. See appendix~\ref{Appendix} for details.
\murata{We will mainly focus on the homogeneous perturbation: $\bm{k}=0$, while inhomogeneous perturbations with $\bm{k}\neq 0$ will be addressed only in Figures~\ref{QNMOrbits1210}.}
Figures~\ref{QMMD}-\subref{QMMR} show the QNM spectra for $\theta=0$ (Dirichlet), $\pi$ (Neumann), and $7\pi/4$ (Robin). From these figures, we can see that the \kozuka{QNM spectra depend} on the Robin parameter $\theta$ and, especially, an unstable mode appears for $\theta=7\pi/4$.

\begin{figure}[t]
    \centering
    \includegraphics[scale=1]{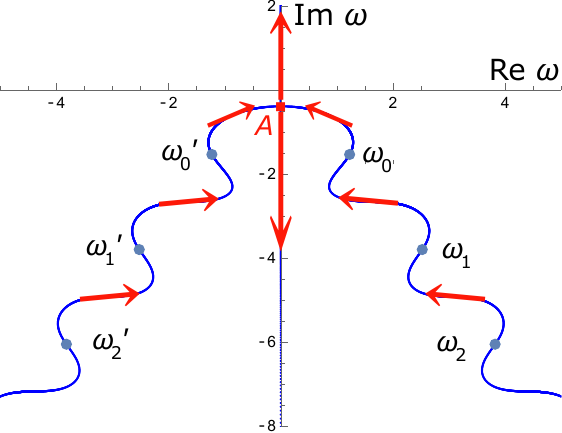}
    \caption{The trajectory of quasinormal frequencies with respect to the Robin parameter $\theta$.}
    \label{QNMOrbits}
\end{figure}

Figure~\ref{QNMOrbits} shows the trajectory of the QNM frequencies in the complex plane 
for one cycle of the Robin parameter $\theta=0\to 2\pi$.
The blue points show the QNM spectrum for the Dirichlet condition $\theta=0$. 
Let $n$ denote the tone number and its QNM frequency be $\omega_n$ (Re $\omega_n \geq 0)$. 
Note that $\omega'_n\equiv -\omega_n^\ast$ is also a QNM frequency. 
First, we track the change of the fundamental frequencies $\omega_0$ and $\omega'_0$ for $\theta=0\to 2\pi$.
As the value of $\theta$ is gradually increasing from $0$, both $\omega_0$ and $\omega'_0$ approach the imaginary axis. 
At $\theta\simeq 1.18\pi$, they merge at the point A on the imaginary axis. 
As $\theta$ is further increasing, they split into two modes on the imaginary axis, and in the limit of $\theta\to2\pi$, each mode asymptotically approaches $\pm \infty i$. 
On the other hand, tracking overtones $\omega_n$ and $\omega'_n$ ($n\geq 1$), we find that they move to the positions of the QNM frequencies with one lower overtone number: 
$\omega_n\to\omega_{n-1}$ and $\omega_n'\to\omega'_{n-1}$ for $\theta=0\to 2\pi$.
After one cycle $\theta=0\to 2\pi$, the distribution of the QNM spectrum returns to its original one but, if we focus on tracking each QNM frequency, it never returns to its original position.
In this sense, there is holonomy in the parametric cycle $\theta=0\to 2\pi$. 
\kozuka{ When we focus on a set of eigenvalue and eigenfunction, this set will not return to the initial set but change into another one after a cycle of continuous variation of an external parameter. Such phenomena have been also reported in other systems and called an eigenvalue holonomy, an exotic holonomy, and so on (for example, see \cite{Cheon:1998pt,Cheon:2008nu,Tanaka:2009zza} and references therein).}

\kozuka{
Figures~\ref{QNMOrbits1}-\subref{QNMOrbits10} show QNM spectra for several wavenumbers $k=1,2,10$.
It can be confirmed that the overall shape of the QNM spectrum remains unchanged even when varying the value of $k$. By increasing the value of $k$, the trajectory of the QNM approaches the real axis but does not intersect with it until fundamental tones merge on the imaginary axis.
}

\kozuka{Figure~\ref{QNMTheta} shows the imaginary parts of QNM frequencies as the function of the Robin parameter $\theta$. The curves in this figure are for $\omega_0,\omega_1$ and $\omega_2$.
Point A in this figure is identical to that in figure~\ref{QNMOrbits}-\ref{QNMTheta}.  Fundamental tones merge at this point and split into to two mode with pure imaginary frequencies. This is the reason why the curve for $\omega_0$ bifurcates into two after the point A.
\begin{figure}[t]
  \centering
  \includegraphics[scale=0.3]{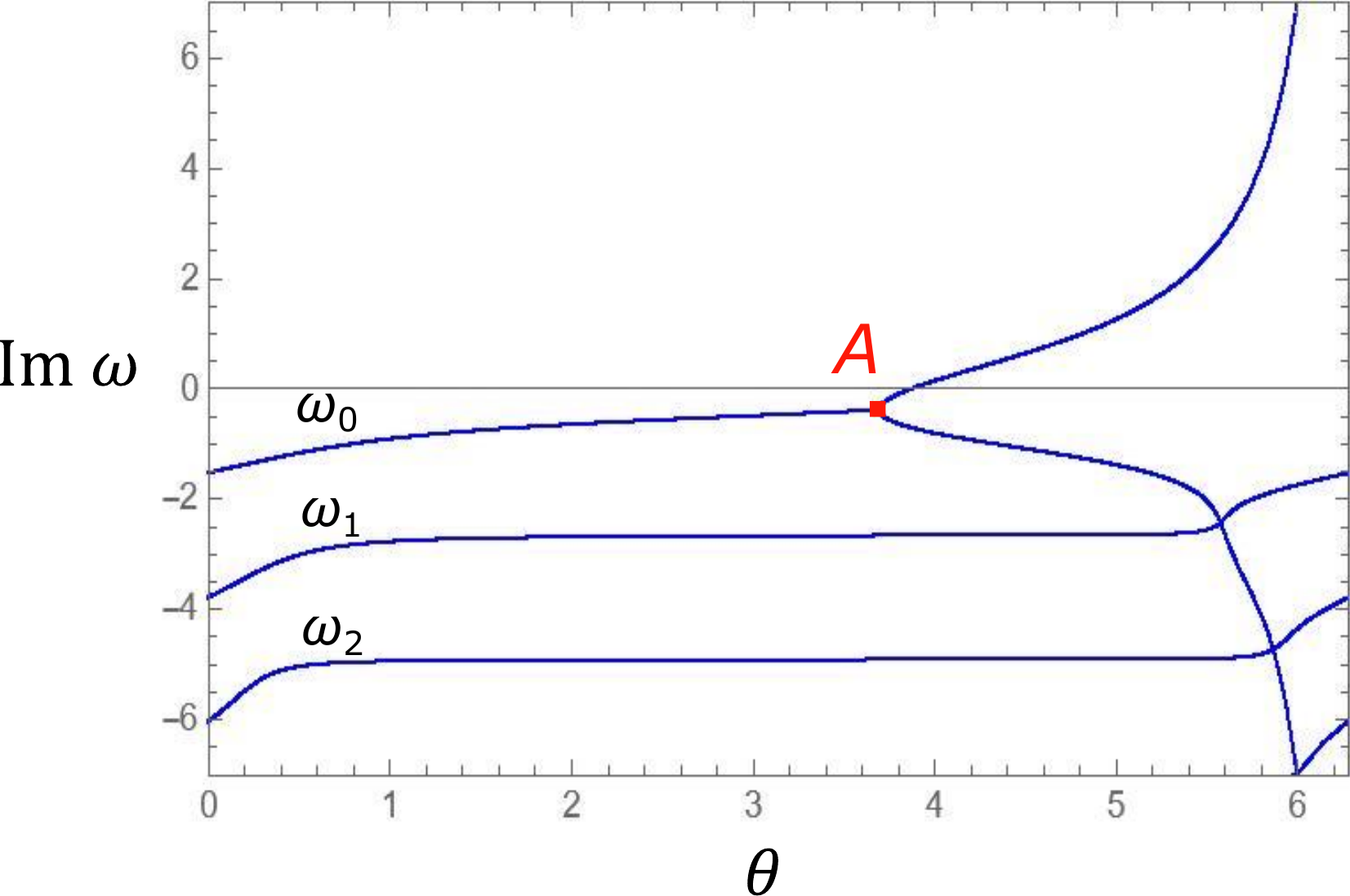}
  \caption{$\theta$ dependence of the imaginary part of the QNM frequencies.}
  \label{QNMTheta}
\end{figure}}

\begin{figure}
  \centering
\subfigure[Re$f(z)$]
 {\includegraphics[scale=0.21]{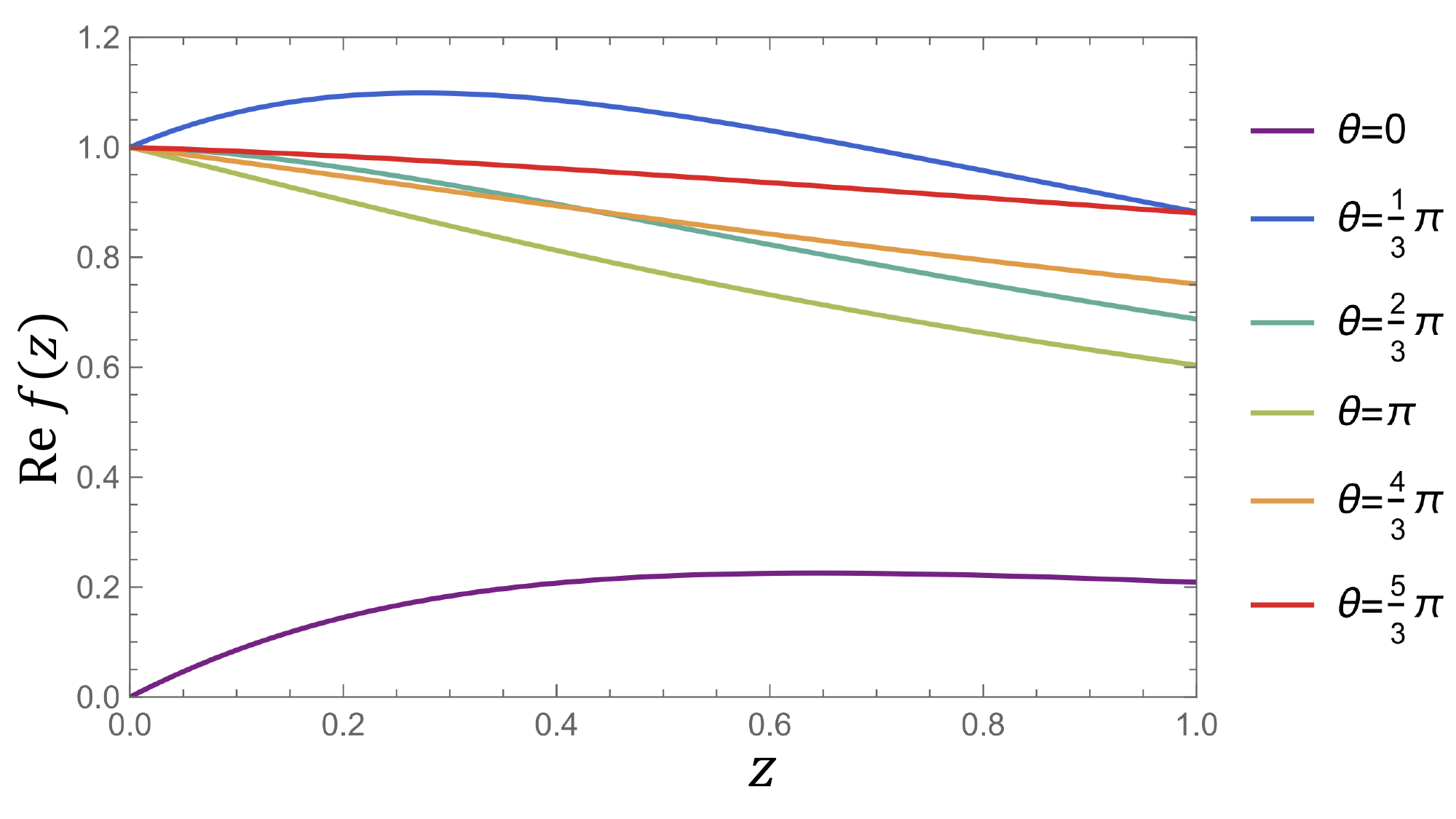}\label{Refz}
  }
  \subfigure[Im$f(z)$]
 {\includegraphics[scale=0.21]{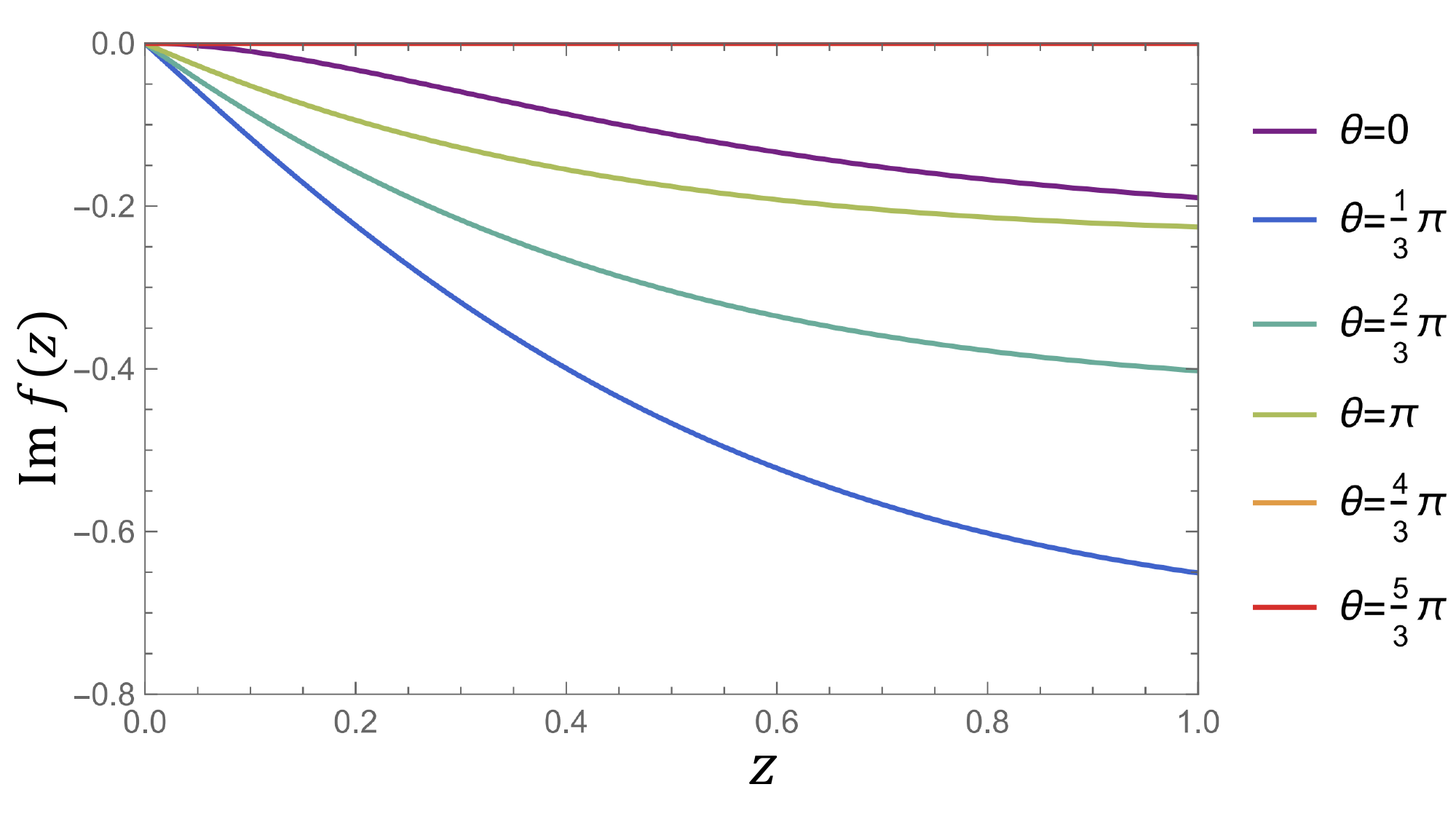}\label{Imfz}
  }
 \caption{
Mode functions of the fundamental tone for $\theta=\pi m/3$, ($m=0,1,2,\cdots 5$).
 }
 \label{QNMmodefunc}
\end{figure}

Figure~\ref{QNMmodefunc} shows the real and imaginary parts of mode functions with the fundamental tone for various Robin parameters $\theta=m\pi/3$ ($m=0,1,\ldots,5$).
We have normalized the mode functions as $f(0)=1$ and $f'(0)=1$ for $\theta\neq 0$ and $\theta=0$, respectively.
Note that the mode function becomes a real function when $\omega$ is pure imaginary.

As in Figure~\ref{QNMOrbits}, we find that the QNM frequencies of the fundamental tones have swept out to infinity as $\omega\to \pm \infty i$ $(\lambda \to \pm \infty)$ in the limit of $\theta\to 2\pi-0$. Thus, for $\theta\simeq 2\pi-0$, supposing $|\lambda| \gg 1$, the principal part of differential equation\ \eqref{eqn0} can be approximately written as $\lambda f'(z)\simeq 0$, and then its solution is given by $f(z)=\textrm{const.}$
In fact, the mode function approaches a constant as $\theta\to 2\pi-0$ as shown in Fig.~\ref{Refz}.

\subsection{Onset of the instability}

As one can see in Figure~\ref{QNMTheta}, the unstable mode appears depending on the Robin parameter $\theta$.
 \kozuka{
 As in Fig.~\ref{QNMOrbits}, we have $\omega=0$ at the onset of instability, i.e, the perturbation becomes non-dynamical.  
 We can show this fact from a general argument. At the onset of instability, we have $\omega\in \mathbb{R}$ because the imaginary part of $\omega$ should be zero. Then, 
 from Eq.~(\ref{Schro}), the Wronskian $W=\chi^\ast d\chi/dr_\ast - \chi d\chi^\ast/dr_\ast$ is conserved along the radial direction: $dW/d r_\ast=0$. Near the horizon, the scalar field behaves as $\chi(z)\propto e^{-i\omega r_\ast}$ and the Wronskian becomes $W|_{z=1}\propto \omega$. On the other hand, near the AdS boundary, we have $\chi\simeq \phi_1 + \phi_2 z$ and $W|_{z=0}=0$. 
 Therefore, the perturbation must be non-dynamical ($\omega=0$) at the onset. 
 }
 
\begin{figure}[t]
    \centering
    \includegraphics[scale=0.3]{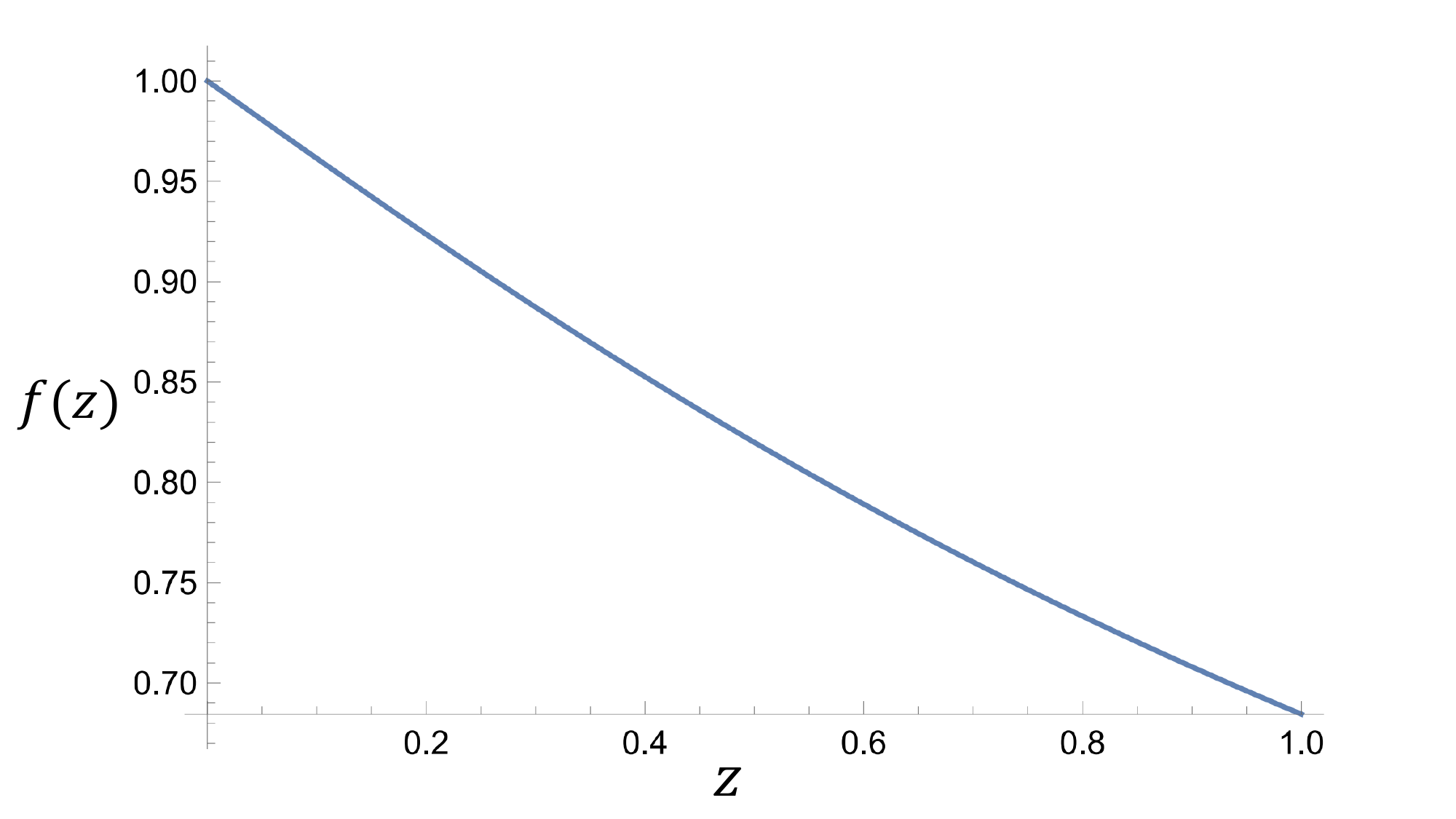}
    \caption{The mode function at the onset of the instability.
    }
    \label{regularfzgraph}
\end{figure}

To consider the onset of instability, we substitute \kozuka{$\lambda=0$ ($\omega=0$)} into Eq.\ \eqref{eqn0} and solve the second-order differential equation. The general solution is given by
\begin{equation}
f(z)= c_1\ _2F_1\qty(\frac{1}{3},\frac{1}{3};\frac{2}{3};z^3)+c_2z\ _2F_1\qty(\frac{2}{3},\frac{2}{3};\frac{4}{3};z^3),
\end{equation}
where $_2F_1(a,b;c;z)$ is the Gaussian hypergeometric function.
Each of the hypergeometric functions diverges logarithmically at $z\rightarrow1$. In order for the solution to be regular at $z=1$, 
we will choose the coefficients $c_1$ and $c_2$ so as to cancel out the terms of $\log(1-z)$ in the two hypergeometric functions.
As the result, we have
\begin{equation}
\begin{aligned}
    f(z)=f_0(z) &\equiv{}_2F_1\qty(\frac{1}{3},\frac{1}{3};\frac{2}{3};z^3)-\frac{3\Gamma(2/3)^3}{\Gamma(1/3)^3}z \, {}_2F_1\qty(\frac{2}{3},\frac{2}{3};\frac{4}{3};z^3) \\
    &= \frac{\Gamma(2/3)^2}{\Gamma(1/3)} {}_2F_1\qty(\frac{1}{3},\frac{1}{3};1;1-z^3) ,
    \label{regularfz}
\end{aligned}
\end{equation}
where we have chosen an overall coefficient as $f(0)=1$.
The above function is shown in Fig.~\ref{regularfzgraph}.
From Eq.~(\ref{regularfz}), the Robin parameter at the onset of instability can be read out by \eqref{boundary condition} as  
\begin{equation}
    \kappa=\frac{f'_0(0)}{f_0(0)}=-\frac{3\Gamma(2/3)^3}{\Gamma(1/3)^3}\simeq -0.3874 .
    \label{onsetkappa}
\end{equation}
Also, 
the value of $\theta = 2\, \textrm{arccot} \, \kappa$ at the onset of instability is given by $\theta\simeq 1.235\pi\simeq 222.4^\circ$. 
Therefore, the range of $\theta$ where the instability occurs is
\begin{equation}
    222.4^\circ \simeq 2\,\textrm{arccot}\left(-\frac{3\Gamma(2/3)^3}{\Gamma(1/3)^3}\right)\leq\theta<360^\circ .
    \label{unstabletheta}
\end{equation}
At the onset of instability~(\ref{onsetkappa}), we can find that $\Phi(t,z)=tzf_0(z)$ is an exact solution of the Klein-Gordon equation.
It follows that the Sch-AdS$_4$ is marginally unstable at the onset.

\subsection{Origin of instability}
\label{origininst}

In this subsection we argue origin of instability under the Robin boundary condition in asymptotically AdS spacetime as in Sec.~\ref{subsec:origin_flat}. 
The renormalized action for a scalar field \cite{deHaro:2000vlm} is given by 
\begin{equation}
    S[\Phi] = - \frac{1}{2}\int_{z \ge \epsilon} d^4 x \sqrt{-g} \left(g^{\mu\nu}\nabla_\mu\Phi\nabla_\nu\Phi + \mu^2 \Phi^2\right)
    - \frac{\Delta_-}{2 l}\left.\int d^3x \sqrt{-h} \Phi^2\right|_{z=\epsilon} ,
    \label{eq:Sren}
\end{equation}
where $h_{\mu\nu}$ denotes the induced metric on the AdS boundary $z=\epsilon$ for $\epsilon \to 0$. 
Note that the last term is a counter term for the action to regularize at the AdS boundary.
In terms of $\Phi \to \Phi + \delta\Phi$, the variation of the action with imposing the field equation is 
\begin{equation}
\begin{aligned}
    \delta S &= \left.\int d^3 x (\sqrt{-g} g^{zz}\Phi' - l^{-1}\Delta_- \sqrt{-h}\Phi) \delta\Phi \right|_{z=\epsilon}\\
    &= \left. \int d^3 x  \sqrt{-h}(\sqrt{g^{zz}}\Phi' - l^{-1}\Delta_-\Phi) \delta\Phi \right|_{z=\epsilon}\\
    &= l^2(\Delta_+ - \Delta_-)\left.\int d^3 x \phi_2 \delta\phi_1 \right|_{z=\epsilon} .
\end{aligned}
\end{equation}
Thus, the on-shell action depends on the boundary data of the scalar field $\phi_1$.
To make the variation of the action vanish under the Robin boundary condition $\phi_2 = \kappa \phi_1$, 
we should modify the action as 
\begin{equation}
    \widetilde{S} = S - l^{2}(\Delta_+ - \Delta_-)\int d^3 x \frac{\kappa}{2} \phi_1^2 .
\end{equation}

The Hamiltonian derived from the action (\ref{eq:Sren}) is  
\begin{equation}
    \begin{aligned}
        H &= \int dx^2 \int dz \sqrt{-g} (-T^0{}_0) + \frac{\Delta_-}{2 l}\left.\int d^2x \sqrt{-h} \Phi^2\right|_{z=\epsilon} ,
    \end{aligned}
\end{equation}
where $T_{\mu\nu}$ is the minimal stress tensor for the scalar field given by 
\begin{equation}
    T_{\mu\nu} = \nabla_\mu\Phi \nabla_\nu\Phi - \frac{1}{2}g_{\mu\nu} \left(\nabla^\rho\Phi\nabla_\rho\Phi + \mu^2\Phi^2\right) .
\end{equation}
This is equivalent to the Killing energy with respect to the time-translational Killing vector $\xi^\mu$.
The local conservation law for the stress tensor, $\nabla_\mu (T^{\mu\nu} \xi_\nu)=0$, yields  
\begin{equation}
\begin{aligned}
    \frac{d}{dt}H &= \int d^2x\int_0^{z_h} dz \frac{\partial}{\partial t}(-\sqrt{-g}T^0{}_0) 
    + \frac{\Delta_-}{l}\left.\int d^2x \sqrt{-h} \Phi \dot\Phi\right|_{z=\epsilon} \\
    &= \left.\int d^2x \left(- \sqrt{-g} T^z{}_0 + \frac{\Delta_-}{l} \sqrt{-h} \Phi \dot\Phi\right)\right|_{z=0} \\
    &= -\left.\int d^2x \sqrt{-h}\left(\sqrt{g^{zz}}\Phi' - \frac{\Delta_-}{l} \Phi\right)\dot\Phi\right|_{z=0} \\
    &= -l^2 (\Delta_+ - \Delta_-) \int d^2x \phi_2\dot\phi_1 .
\end{aligned}
\end{equation}
It turns out that the Hamiltonian is conserved if the Dirichlet condition $\phi_1 = 0$ or the \kozuka{Neumann condition $\phi_2 = 0$ is satisfied}, while it cannot be conserved otherwise.  
Indeed, when we impose the Robin boundary condition $\phi_2=\kappa \phi_1$, the energy flux appears at the AdS boundary.

Now, we define the modified energy, corresponding to the Noether charge derived from the modified action $\widetilde{S}$, as 
\begin{equation}
\begin{aligned}
    E_\text{mod} &\equiv H + l (\Delta_+ - \Delta_-) \frac{\kappa}{2} \int d^2x \phi_1^2 \\
    &= \int dx^2 \int dz \sqrt{-g} (-T^0{}_0) + \frac{\Delta_-}{2 l}\left.\int d^2x \sqrt{-h} \Phi^2\right|_{z=\epsilon}\\
    & \quad + l (\Delta_+ - \Delta_-) \frac{\kappa}{2} \int d^2x \phi_1^2 ,
\end{aligned}
\end{equation}
which can be conserved at the AdS boundary under the Robin boundary condition $\phi_2 = \kappa \phi_1$.
Since $\Delta_+ - \Delta_- = \sqrt{9 +\mu^2} > 0$, if $\kappa <0$, $\Phi$ could be unbounded with the modified energy kept conserved.
As in the case of the $(1+1)$-wave equation, the Robin boundary condition with negative $\kappa$ works as if infinite energy is stored at the AdS boundary.

It is worth noting that the modified action and energy discussed above is related to the double-trace deformation of the boundary CFT in the context of the AdS/CFT correspondence.
In order for the variation problem of the bulk action to be well-defined, the action is chosen so that the contribution of the boundary term should vanish under the Robin boundary condition for the bulk field.
Then, the energy defined by the Noether charge for that action will be conserved and such action and energy are identical to the ones discussed here on shell.

\section{Conclusion}\label{conc}
We investigated the QNM spectrum of scalar fields on Sch-AdS$_4$ imposing the Robin boundary condition. As a result, as can be seen from Figure~\ref{QNMOrbits} and Figure~\ref{QNMTheta}, after the parametric cycle of the Robin parameter, $\theta=0\to 2\pi$, while the spectrum itself does not change, the fundamental tone is swept out to the infinity in the complex plane and the $n$-th overtone moves to the $(n-1)$-th overtone for $n\geq 1$. In this sense, there is holonomy under the cycle $\theta=0\to 2\pi$. We also found the unstable mode for the scalar field perturbation  under the Robin boundary condition with $222.4^\circ\leq \theta < 360^\circ$. Furthermore, we numerically obtained the $\theta$ dependence of the mode function of the fundamental tone. 

This instability arises from the Robin boundary condition with negative $\kappa$. The origin can be intuitively understood as follows. The Robin boundary condition with negative $\kappa$ corresponds to a delta-function potential well and can work as an infinite energy storage. Indeed, under the Robin boundary conditions, the conserved energy as Noether charge is modified by adding a boundary contribution, whose sign depends on the Robin parameter $\kappa$. As a result, when $\kappa$ is negative, the ordinary bulk energy of the scalar field could be unbounded while the modified energy remains constant. We can expect that this type of instability commonly exists for probe fields in asymptotically AdS spacetimes.

For $\theta=2\pi$ or $\theta=0$, the boundary condition is just Dirichlet and there is no instability. On the other hand, for $\theta=2\pi -\epsilon$ ($0<\epsilon\ll 1$), which corresponds to a Robin boundary condition with an extremely large negative $\kappa$, the growth rate of instability has an extremely large value, whereas the instability does not occur for $\theta=\epsilon$. (In terms of delta-function potentials, the Robin boundary condition for $\theta = 2\pi -\epsilon$ corresponds to a very deep potential well and that for $\theta=\epsilon$ to a very high potential barrier. Thus, in real time dynamics of the scalar field, we might be able to think that the unstable mode dominates in an instant for $\theta=2\pi -\epsilon$. Does the tiny change of the boundary condition significantly change the real time dynamics of the scalar field? The study of the real time dynamics of the scalar field with the Robin boundary condition is an interesting future direction. We might have to take into account the ``excitation factor''~\cite{Berti:2006wq} to understand dynamics of the scalar field under the Robin boundary condition.

\kozuka{ In this paper we have set $\mu^2 = -2$ in order to avoid numerical difficulties in imposing the boundary condition on functions with irrational power-law behavior.
We can expect that the QNM spectrum smoothly changes even when $\mu^2$ is varied and 
qualitative behaviors such as holonomy will remain unchanged.}

We found the instability of the Sch-AdS$_4$ under the scalar field perturbation with the Robin boundary condition. At the onset of the instability, there is the static perturbation as in Eq.~(\ref{regularfz}). We can expect that there is a family of nonlinear solutions branching off from the onset.  
Explicit construction of such hairy black hole solutions would be the other future direction.

\section*{Acknowledgments}
The authors would like to thank Akihiro Ishibashi and Takaaki Ishii for useful discussions.
The work of K.~M.~was supported in part by JSPS KAKENHI Grant No.~JP20K03976, JP21H05186 and JP22H01217.

\appendix 
\section{Detail of the numerical calculation of the QNM}\label{Appendix}

We describe the detail of the numerical method to compute the QNM spectrum.
This method is based on the Yaffe's Mathematica notebook in the Mathematica Summer School on Theoretical Physics~\cite{Yaffe}. In the notebook, the \kozuka{pseudospectral method} is used to solve the eigenvalue problem for differential equations.

\begin{figure}
    \centering
 {\includegraphics[scale=0.3]{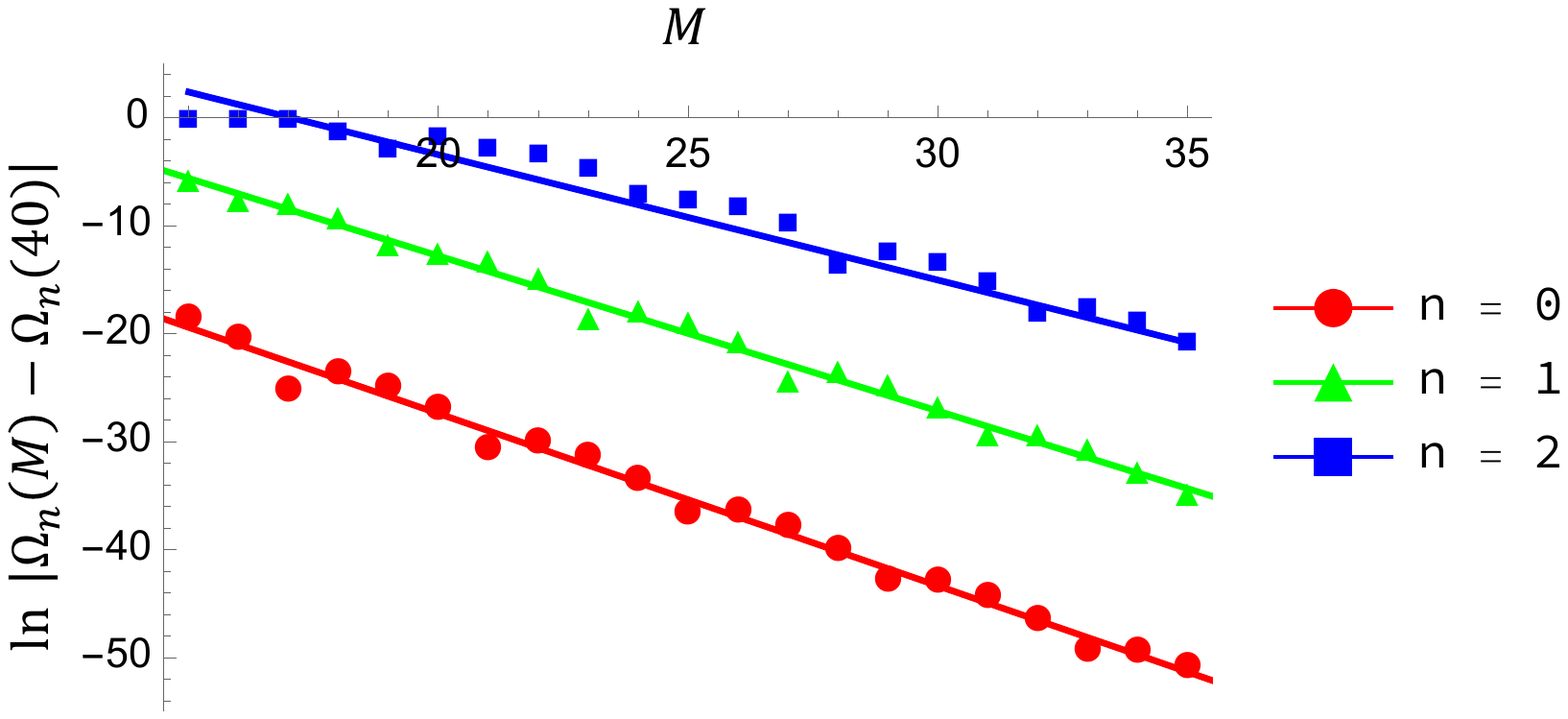}}
 \caption{
\kozuka{The results of convergence test \murata{for tone number $n=0,1,2$}. 
Plot data \murata{$\Omega_n(M)=-\text{Im}~\omega_n(M)$} are fitted for \murata{test functions: $\Omega_0^\text{fit}(M)=\Omega_0(40)-e^{4.5053-1.5952M}$, $\Omega_1^\text{fit}(M)=\Omega_1(40)-e^{15.928-1.4362M}$, $\Omega_2^\text{fit}(M)=\Omega_2(40)-e^{19.784-1.1606M}$}
 by using \texttt{Fit} function. }\murata{As tone number $n$ increases, the decrease in error is gradual.}
 }\label{logfit}
\end{figure}

We introduce the Chebyshev polynomial as
\begin{equation}
    T_n(x)\equiv\cos(n\arccos{x}) \qquad (n=0,1,2,\ldots).
\end{equation}
The Chebyshev polynomials form a complete orthogonal system of functions defined in $-1\leq x\leq1$. We expand a mode function $f(z)$ by the Chebyshev polynomials. 
In actual calculations, we truncate the expansion at the $M+1$ order and write it as
\begin{equation}
    f(z)={\sum_{n=0}^{M+1}}C{_n}T{_n}(2z-1),
    \label{fx}
\end{equation}
where $0\leq z \leq 1$.
We define Chebyshev collocation points $z_i$ as
\begin{equation}
    z{_i}=\frac{1}{2}\left(1-\cos(\frac{i}{M}\pi)\right) \qquad(i=0,1,\ldots,M) ,
    \label{zgridi}
\end{equation}
\kozuka{where $M=40$.}
\kozuka{In Eq.~\eqref{zgridi}, intervals of collocation points become smaller around at $z=0, 1$.}
Substituting equation (\ref{fx}) into equation~\eqref{eqn0} and evaluating at $z=z_i$, we obtain a set of   ($M+1$) equations: 
\begin{equation}
    \begin{split}
        \sum^{M+1}_{n=0}\{ [-(k^2+z_i)T_{n}(2z_{i}-1)+(2\lambda-3{z_{i}}^2)(T_{n}(2z_{i}-1))'\\
        +\kozuka{(1-{z_{i}}^3)}(T_{n}(2z_{i}-1))''] {C_{n}}\}=0 \\ (i=0,1,\ldots,M),
    \end{split}
    \label{coeffd}
\end{equation}
where $'\equiv d/dz$.\footnote{
\kozuka{Following~\cite{Yaffe}}, for the quick numerical calculations in Mathematica, we rewrite derivatives of Chebyshev polynomials as 
\[
\frac{d^m}{dx^m}T_{n}(x)=n2^{m-1}(m-1)!{C^{(m)}}_{n-m}(x)
\]
where ${C^{(m)}}_{n}(x)$ \kozuka{are} the Gegenbauer polynomials. 
}
Equations\ \eqref{coeffd} can be written in the matrix form as
\begin{equation}
    \begin{pmatrix}
        a_{00}&\cdots&a_{n0}&\cdots&a_{M+1,0}\\
        a_{01}&\cdots&a_{n1}&\cdots&a_{M+1,1}\\
        \vdots& &\vdots& &\vdots\\
        a_{0i}&\cdots&a_{ni}&\cdots&a_{M+1,i}\\
        \vdots& &\vdots& &\vdots\\
        a_{0M}&\cdots&a_{nM}&\cdots&a_{M+1,M}
    \end{pmatrix}
    \begin{pmatrix}
        C_{0}\\
        C_{1}\\
        \vdots\\
        C_{n}\\
        \vdots\\
        C_{M+1}
    \end{pmatrix}
    =0.
    \label{adeterminant}
\end{equation}
The above matrix is not square but $(M+1)\times (M+2)$.
We obtain the other equation from the Robin boundary condition at $z=0$. 
\kozuka{ Although we should impose the ingoing boundary condition at the horizon $z=1$, in the present formulation it corresponds to requiring regularity. Since the solution has been expanded in terms of regular functions at $z=1$, we do not need any condition at $z=1$, additionally. }
Plugging Eq.\ \eqref{fx} into Eq.~\eqref{boundary condition}, we have
\begin{equation}
    \left(\sum^{M+1}_{n=0}C_{n}T_{n}(2z-1)\right)'\bigg|_{z=0} +(\lambda-\kappa)\left(\sum^{M+1}_{n=0}C_{n}T_{n}(2z-1)\right)\bigg|_{z=0}=0 .
    \label{b}
\end{equation}
The above equation is written as
\begin{equation}
    \begin{pmatrix}
        b_{0}&\cdots&b_{n}&\cdots&b_{M+1}
    \end{pmatrix}
    \begin{pmatrix}
        C_{0}\\
        \vdots\\
        C_{n}\\
        \vdots\\
        C_{M+1}
    \end{pmatrix}
    =0 .
    \label{bdeterminant}
\end{equation}
Combining Eq.\ \eqref{bdeterminant} and Eq.\ \eqref{adeterminant}, we obtain
\begin{equation}
    \begin{pmatrix}
        a_{00}&\cdots&a_{n0}&\cdots&a_{M+1,0}\\
        a_{01}&\cdots&a_{n1}&\cdots&a_{M+1,1}\\
        \vdots& &\vdots& &\vdots\\
        a_{0i}&\cdots&a_{ni}&\cdots&a_{M+1,i}\\
        \vdots& &\vdots& &\vdots\\
        a_{0M}&\cdots&a_{nM}&\cdots&a_{M+1,M}\\
        b_{0}&\cdots&b_{n}&\cdots&b_{M+1}
    \end{pmatrix}
    \begin{pmatrix}
        C_{0}\\
        C_{1}\\
        \vdots\\
        C_{n}\\
        \vdots\\
        C_{M}\\
        C_{M+1}
    \end{pmatrix}
    \equiv A\Vec{C}=0 .
    \label{ACmatrix}
\end{equation}
The matrix $A$ is now square. 
Since the components of the coefficient matrix $A$ is linear in the eigenvalue $\lambda=i\omega$,
we can decompose the matrix $A$ into two matrices $\alpha$ and $\beta$ as 
\begin{equation}
    A=\alpha-\lambda\beta ,
\end{equation}
where $\alpha$ and $\beta$ do not depend on $\lambda$.
Then, Eq.\ \eqref{ACmatrix} takes the form of a generalized eigenvalue equation:
\begin{equation}
    \alpha\Vec{C}=\lambda\beta\Vec{C} .
\end{equation}
We obtain the spectrum of QNM as eigenvalues by \kozuka{using \texttt{Eigenvalues} command to solve} this equation in Mathematica numerically.
In the actual computation, we adopted $M=40$ and checked that results shown in this paper do not change much even if we increase $M$ to $80$ \kozuka{since they converge exponentially. The results of convergence test is shown in Fig.~\ref{logfit}.}
\murata{Since the truncation errors depending on the cutoff $M$ seem to sufficiently converge to 
the machine precision for $M\gtrsim 40$, we only showed plots for $M\lesssim 40$ in Fig.~\ref{logfit}.}

\bibliography{robinqnm}

\end{document}